\documentclass[aps,amsfonts,nofootinbib,superscriptaddress,twocolumn]{revtex4-1}

\usepackage{appendix}
\usepackage{dsfont}
\usepackage{amssymb}
\usepackage{amsmath}
\usepackage{amsbsy}
\usepackage{color}
\usepackage{slashed}
\usepackage{bm}
\usepackage{bbm}
\usepackage{dutchcal}
\usepackage{tikz}
\usepackage{xparse}
\usepackage[normalem]{ulem}

\usetikzlibrary{calc}
\usetikzlibrary{positioning,arrows,patterns,automata}
\usetikzlibrary{decorations.pathmorphing,decorations.markings,trees}
\usetikzlibrary{shapes.misc}

\tikzset{	photon/.style={decorate, decoration={snake}, draw=black},
		particle/.style={draw=black, postaction={decorate},
        			decoration={markings,mark=at position .5 with {\arrow[draw=black]{>}}}},
	    	aparticle/.style={draw=black, postaction={decorate},
        			decoration={markings,mark=at position .5 with {\arrow[draw=black]{<}}}},
	    	axion/.style={draw=black},
		gluon/.style={decorate, draw=red,
        			decoration={coil,amplitude=4pt, segment length=5pt}},
		vertex/.style={draw,shape=circle,fill=black,minimum size=80pt,inner sep=1pt},
		cross/.style={cross out, draw=black, minimum size=2*(#1-\pgflinewidth), inner sep=0pt, outer sep=0pt}, 				cross/.default={0.3cm}}
		
 \NewDocumentCommand\semiloop{O{black}mmmO{}O{above}}
{\draw[#1] let \p1 = ($(#3)-(#2)$) in (#3) arc (#4:({#4+180}):({0.5*veclen(\x1,\y1)})node[midway, #6] {#5};)}

\newcommand{\ee}{\end{equation}}
\newcommand{\bb}{\begin{equation}}
\newcommand{\eqb}{\begin{eqnarray}}
\newcommand{\eqf}{\end{eqnarray}}

\begin{document}
\title{ Anapole Dark Matter Interactions as Soft Hidden Photons}
 \author{
 Paola Arias 
 }
\email{paola.arias.r@usach.cl}
\affiliation{Departamento de  F\'{\i}sica, Universidad de  Santiago de
  Chile, Casilla 307, Santiago, Chile}
  \author{
 J. Gamboa 
 }
\email{jorge.gamboa@usach.cl}
\affiliation{Departamento de  F\'{\i}sica, Universidad de  Santiago de
  Chile, Casilla 307, Santiago, Chile}
   \author{
Natalia Tapia 
 }
\email{natalia.tapiaa@usach.cl}
\affiliation{Departamento de  F\'{\i}sica, Universidad de  Santiago de
  Chile, Casilla 307, Santiago, Chile} \date{\today}
\begin{abstract} 
We propose a model where the anapole appears as a hidden photon that is coupled to visible matter through a kinetic mixing. 
For low momentum $|{\bf p}| \ll M$ where $M$ is the cutoff the model (soft hidden photon limit) is reduced to the Ho-Scherrer description.  We show that the hidden gauge boson is stable and therefore hidden photons are indeed, candidates for dark matter. Our approach shows that anapole and kinetic mixing terms are equivalent descriptions seen from different scales of energy.\end{abstract}
\date{\today}
\maketitle


Majorana fermions are particles that feature good attributes to be considered dark matter \cite{DM} candidates. On the one hand they are electrically neutral and can interact with virtual photons. However, there are several ways to implement the  dark matter idea as,  for example, by including  neutralinos \cite{neutralinos}, analyzing relic abundance systematically \cite{relic}, incorporating Sommerfeld enhancement  \cite{arkani,otros} or extending the standard model through the use of secret interactions \cite{secret}, kinetic mixing \cite{kinetic} and so on. 

However there are also other reasons that justify an additional study of Majorana fermions as dark matter, namely, it is expected that the effects of dark matter will be more accessible in the low energy sector, and if the dark matter is coupled with the standard model, then  effects such as parity violation  can play an important role \cite{kamion}.
\skip 0.5cm

As in any effective theory one expects that there is a cutoff energy $M$, for which if $ E \simeq M $, both the parity violation and the anapole contributions provide visible signals   as a consequence, most likely, of more fundamental symmetries unknown until now.

The consequences of the description above are followed by the fact that the vertex function $\Gamma^\mu (q^2)$  must be consistent with gauge invariance because it is related to the electromagnetic current through
\[ 
\langle k| J^\mu | k' \rangle = {\bar u}({\bf k}) \Gamma^\mu (q^2) u({\bf k}'), 
\] 
where $\Gamma^\mu (q^2)$ has the following general structure
\eqb
\Gamma^\mu &=&  F_1 (q^2)\gamma^\mu + F_2(q^2) \frac{i}{2{\bar m}} \sigma^{\mu \nu} p_\nu  \nonumber 
\\ 
&+& F_3 (q^2) \frac{i}{2m} \sigma^{\mu \nu} q_\nu \gamma_5  + F_4 (q^2) (\gamma^\mu  q^2 -q^\mu {{ q }  \hspace{-.5em}  \slash\hspace{.25em}}) \gamma_5, 
\eqf
where the $ F_{1,2,3,4} $ are form factors and $q^2 =-(k-k')^2$. 

The term proportional to $F_3$ is the electric dipole moment which violates temporal inversion but is invariant under parity, while the term proportional to $ F_4 $ is called the anapole contribution \cite{zel} and violates both parity and temporal inversion.

The Lagrangian that provides the anapole contribution is \cite{scherrer,mas}
\bb
{\cal L}_{\mbox{anapole}}= -\frac{g}{M^2} {\bar \chi} \gamma_\mu \gamma_5 \chi ~\partial_\nu F^{\mu \nu}, \label{anaa1}
\ee
where $F_{\mu \nu} = \partial_\mu A_\nu -\partial_\nu A_\mu$ is the strength tensor, $g$ is a dimensionless coupling constant and $M$ is a cutoff of mass. 

In a more classical context, and inspired by the remarkable ideas developed in the late fifties in weak interactions physics \cite{yang-lee}, the hypothesis of identifying Majorana fermions with dark matter is even more intriguing because the electromagnetic interaction \cite{zel} with Majorana fermions should occur through the anapole term \cite{scherrer,mas}. 

Intuitively, the anapole interaction appears when neutral fermions are coupled through the process shown in  FIG. 1, where the black box vertex encodes the parity violation \cite{zel}. Then, the possible relation between dark matter and fermions of the standard model ($f$) emerges when we \lq \lq paste" the processes 
$\chi {\bar \chi} \rightarrow \gamma $ and $ \gamma \rightarrow {\bar f} f $, with  $ \gamma $ a virtual photon playing the role of a \lq \lq bridge" between dark matter and the standard model (see FIG. 2).
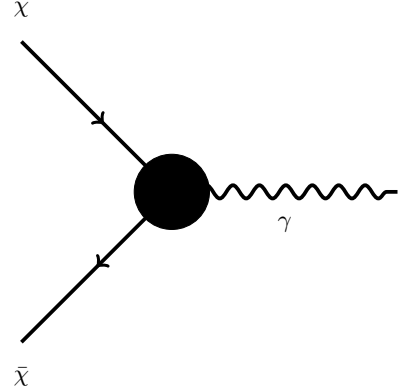
\begin{figure}
\begin{tikzpicture}[node distance=1cm and 1cm]
\coordinate (O) at (0, 0);
\draw[line width=0.5mm, aparticle] (O)--(-2,2) node[left, above=0.2cm] {$\chi$};
\draw[line width=0.5mm, particle] (O) -- (-2,-2) node[left, below=0.2cm] {$\bar{\chi}$};
\draw[line width=0.5mm, photon] (O) -- (3,0)node[midway, below=0.2cm] {$\gamma$};
\filldraw[fill=black, draw=black] (O) circle (0.5cm);
\end{tikzpicture}
\caption{Annihilation  of two Majorana fermions into a virtual photon. The black box is an effective vertex.}
\end{figure}

The goal of this paper is to  present an approach that shows a very clear relationship between hidden photons and anapole as dark matter. In a nutshell, our results show that the anapole term can emerge from a  soft hidden photon limit approximation.

An extra U(1) gauge boson -  often referred as hidden photon or dark photon - it is a quite interesting and well motivated dark matter candidate. Has been pointed out that such particle can explain the whole  dark matter content observed today, or be a sub-dominant component, if produced a non-thermally during inflation \cite{Graham:2015rva}. 
\begin{figure}
\begin{minipage}{0.5\textwidth}
\begin{tikzpicture}[node distance=1cm and 1cm]
\coordinate (O) at (0, 0);
\draw[line width=0.5mm, aparticle] (O)--(-2,2) node[left, above=0.2cm] {$\chi$};
\draw[line width=0.5mm, particle] (O) -- (-2,-2) node[left, below=0.2cm] {$\bar{\chi}$};
\draw[line width=0.5mm, photon] (O) -- (1.5,0);
\draw[line width=0.5mm, photon] (2,0) -- (3.5,0);
\filldraw[fill=black, draw=black] (O) circle (0.5cm);
\draw[line width=0.5mm,aparticle] (3.5,0)--(5.5,2) node[left, above=0.2cm] {$SM$};
\draw[line width=0.5mm, particle] (3.5,0)--(5.5,-2) node[left, below=0.2cm] {$SM$};
\filldraw[fill=white, draw=white] (3.5,0) circle (0.5cm);
\filldraw[fill=white, draw=black,pattern=north west lines, pattern color=black] (3.5,0) circle (0.5cm);
\draw[line width=0.5mm, dashed] (1.75,-2)--(1.75,2) ;
\end{tikzpicture}
\end{minipage}
\begin{minipage}{0.5\textwidth}
\begin{tikzpicture}[node distance=1cm and 1cm]\centering
\coordinate (O) at (0, 0);
\draw[line width=0.5mm, aparticle] (O)--(-2,2) node[left, above=0.2cm] {$\chi$};
\draw[line width=0.5mm, particle] (O) -- (-2,-2) node[left, below=0.2cm] {$\bar{\chi}$};
\draw[line width=0.5mm, photon] (O) -- (3,0);
\filldraw[fill=black, draw=black] (O) circle (0.5cm);
\draw[line width=0.5mm,aparticle] (3,0)--(5,2) node[left, above=0.2cm] {$SM$};
\draw[line width=0.5mm, particle] (3,0)--(5,-2) node[left, below=0.2cm] {$SM$};
\draw[line width=0.5mm, photon] (O) -- (3,0)node[midway, below=0.2cm] {$\gamma$};
\filldraw[fill=white, draw=white] (3,0) circle (0.5cm);
\filldraw[fill=white, draw=black,pattern=north west lines, pattern color=black] (3,0) circle (0.5cm);
\end{tikzpicture}
\end{minipage}
\caption{This diagram `pastes' the dark (LHS) and visible (RHS) parts through a virtual photon.}
\end{figure}
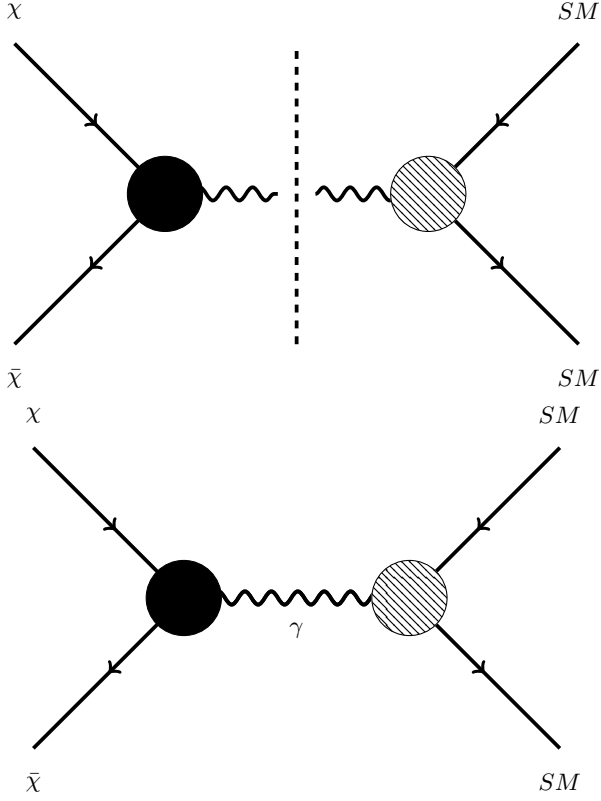
 
 
 In order to explain the idea we begin considering the Lagrangian
 \bb
 {\cal L}= {\cal L} (\{\chi,G_\mu\}, \{\psi, G_\mu\}), \label{xx}
 \ee
  where $\{\chi, G_\mu\}$ correspond to the particle content of the dark sector, with $G_\mu$  a hidden $U(1)$ gauge field, whereas in $\{\psi, A_\mu \} $ the
$\psi$'s are fermions of the standard model and $A_\mu$ is the visible photon. 
  
  More specifically (\ref{xx}) will be written as  
\eqb
{\cal L} &=& {\bar \chi} \left( i {   {{ \partial }  \hspace{-.6em}  \slash
      \hspace{.15em}}} -g  \gamma_5  {   {{ G }  \hspace{-.6em}  \slash
      \hspace{.15em}}}    -{\bar m} \right) \chi  -\frac{1}{4} G_{\mu \nu}^2 (G) +\frac{1}{2} M^2 {G_\mu}^2\nonumber 
      \\
      &+& {\bar \psi} \left( i {   {{ \partial }  \hspace{-.6em}  \slash
      \hspace{.15em}}} -g    {   {{ A }  \hspace{-.6em}  \slash
      \hspace{.15em}}}    -m\right) \psi       
       - \frac{1}{4}F_{\mu \nu}^2(A)  + \frac{\xi}{2} F_{\mu \nu}(A) G^{\mu \nu}(G)  \nonumber 
       \\ 
       && +\cdots , \label{lagra112}
      \eqf
  with the strength  tensors $F_{\mu \nu} (A)$ and $G_{\mu \nu} (G)$ defined as  
\eqb 
F_{\mu \nu} (A)&=& \partial_\mu A_\nu -\partial_\nu A_\mu, \nonumber 
\\ 
G_{\mu \nu} (G)&=& \partial_\mu G_\nu - \partial_\nu G_\mu.  
\eqf 
  
  In (\ref{lagra112}),  $\xi$ parametrizes the (small) kinetic mixing between  hidden and visible photons,  and $\cdots$ denotes all other fields belonging to the standard model. 
  
  A mass term, $M$, for the hidden photon has been included, assuming that there is a particular spontaneous symmetry breaking mechanism in the hidden sector or corresponds to a  stueckelberg mass term (see also \cite{brout}). 
  
  The next step is to consider a region in which the hidden photon momentum satisfies $ |{\bf p}| \ll M $, such limit can be justified for a hidden photon dark matter candidate, since $|{\mathbf{ v}}_{dm}|\ll 1$,  so that in this limit  the kinetic term $G_{\mu \nu}^2$,  is much smaller than $M^2 G_\mu^2$, so  
  \eqb
{\cal L} &=& {\bar \chi} \left( i {   {{ \partial }  \hspace{-.6em}  \slash
      \hspace{.15em}}} -g  \gamma_5  {   {{ G }  \hspace{-.6em}  \slash
      \hspace{.15em}}}    -{\bar m} \right) \chi  +\frac{1}{2} M^2 {G_\mu}^2\nonumber 
      \\
      &+& {\bar \psi} \left( i {   {{ \partial }  \hspace{-.6em}  \slash
      \hspace{.15em}}} -g    {   {{ A }  \hspace{-.6em}  \slash
      \hspace{.15em}}}    -m\right) \psi       
       - \frac{1}{4}F_{\mu \nu}^2(A)  - \xi G_\nu \partial_\mu F^{\mu \nu}(A)  \nonumber 
       \\ 
       && +\cdots. \label{lagra12}
      \eqf
Where  we have made an integration by parts of the   kinetic mixing term. 
   
   In this region of energy, $G_\mu$ becomes an auxiliary field, and therefore can be found to be
   \eqb
   G_\mu &=& \frac{1}{M^2} \left( g{\bar \chi} \gamma_\mu \gamma_5 \chi + \xi \partial^\nu F_{\mu \nu} (A)\right). \nonumber 
   \\
   &=& \frac{1}{M^2} \left( gJ^{(5)}_\mu + \xi \partial^\nu F_{\mu \nu} (A)\right).    \label{aux}
   \eqf
   
   Putting back the latter expression for $G_\mu $ into (\ref {lagra12}) we get
   \eqb
   {\cal L} &\simeq&  {\bar \chi} \left( i {   {{ \partial }  \hspace{-.6em}  \slash
      \hspace{.15em}}}   -{\bar m} \right) \chi  -\frac{g^2}{M^2} \left({\bar \chi} \gamma_\mu \gamma_5 \chi \right)^2  \nonumber 
      \\
      &-&\xi \frac{ g}{M^2}  
       {\bar \chi} \gamma_\mu \gamma_5 \chi  ~ \partial_\nu F^{\mu \nu}  + {\cal O}(\xi^2).      \label{lagra123}
         \eqf
     
  Thus,  from the above equation can be clearly seen that  the anapole term comes from the \lq \lq soft-photons" approximation, meaning from the assumption that the momentum of the  hidden photon is much smaller than $ M $. 
  
  However, we also emphasize that the anapole contribution is  a consequence of the kinetic mixing of the dark matter as it becomes explicit by the presence of $\xi $.  Therefore, a Majorana fermion coupled to the DM particle can manifest itself via an anapole interaction.
  
The measurement of  anapole contributions  is a task that has been developing slowly, where these contributions are considered a test of precision of the standard model. Currently there are several experiments running in atomic physics \cite{flaum}, that complement the first measurement in this direction \cite{wood}.

In the context of  anapole contributions coming from dark matter, these are more difficult to control and what we have proposed in this letter is to address them via kinetic mixing.

The point of view considered here unifies two apparently different approaches and allows to establish a one-to-one correspondence between both.

Another interesting question is to investigate the stability of the $G_\mu$ boson. 
In the static limit this boson is, of course, stable but if $G_\mu$ is a dynamical field, then one should worry about its life time.

In order to do so, it is convenient to take the basis $(A_\mu, G _\mu)$ and diagonalize it, so the mass eigenstates will be
\eqb 
A_\mu && \cong A'_\mu + g \left(1+ \frac{m^2}{M^2}\right) G'_\mu \nonumber 
\\
G_\mu && \cong A'_\mu - g \frac{m^2}{M^2} G'_\mu
\eqf

And the Lagrangian in the hidden sector becomes
\bb 
{\cal L} \subset -g^2 \frac{m^2}{M^2} G'_\mu {\bar \chi}\gamma_5 \gamma^\mu \chi, \nonumber
\ee
so the  decay with is given by \cite{yana}
\bb 
\Gamma \left(G \rightarrow {\bar \chi} \chi  \right) \cong g^2 \left(\frac{m}{M}\right)^4 m.
\ee

However, as in many extensions of the standard model, it is sensible to consider $ m \ll M$, thus, the decay amplitude it is indeed  small,  implying that the average life time of the gauge boson is  large, at least compared to the life time of the universe, so the boson $G_\mu$ is stable. 

From the point of view of the phenomenological possibilities of detection one could explore at least two; the first one requires an exhaustive analysis of the data  XENON100 \cite{xenon}, and the second is to exploit the extra Coulomb contribution that results from  the Lagrangian of eq.~(\ref{lagra12}), namely \cite{flaum}
\[
V(r) = \frac{g}{4 \pi}\frac{e^{-M r}}{r} \gamma_5.
\]
which is the  lowest order contribution.


Classical experiments of Coulomb's law, bounds on the photon mass \cite{pdg} and parity violation tests could, in principle, provide some clues about  the existence of dark matter in the universe.
%
%


We would like to thank Prof. F. Mendez and F. A. Schaposnik by discussions. This work  was  supported  Dicyt  (J.G.) and  Fondecyt-Chile project 
 1161150 (P.A.), and Conicyt Fellowship 21160064 (N.T.).

\end{document}